\documentclass{optica-article}

\journal{boe}

\articletype{Research Article}

\usepackage{lineno}

\begin{document}

\title{3D Super-resolution Optical Fluctuation Imaging with Temporal Focusing two-photon excitation}

\author{Pawel Szczypkowski,\authormark{1,*} Monika Pawlowska,\authormark{1,2} and Radek Lapkiewicz\authormark{1,*}}

\address{\authormark{1}Institute of Experimental Physics, Faculty of Physics, University of Warsaw, Pasteura 5, Warsaw 02-093, Poland\\
\authormark{2}Nencki Institute of Experimental Biology PAS, Pasteura 3, 02-093, Warsaw, Poland\\
}
\email{\authormark{*}Pawel.Szczypkowski@fuw.edu.pl} 
\email{\authormark{*}Radek.Lapkiewicz@fuw.edu.pl}



\begin{abstract}
3D super-resolution fluorescence microscopy typically requires sophisticated setups, sample preparation, or long measurements. A notable exception, SOFI, only requires recording a sequence of frames and no hardware modifications whatsoever but being a wide-field method, it faces problems in thick, dense samples. We combine SOFI with temporal focusing two-photon excitation -- the wide-field method that is capable of excitation of a thin slice in 3D volume. Both methods are easy to implement in a standard microscope, and by merging them, we obtain super-resolved 3D images of neurons stained with quantum dots. Our approach offers reduced bleaching and an improved signal-to-background ratio that can be used when robust resolution improvement is required in thick, dense samples. 
\end{abstract}

\section{Introduction}

Biological processes often occur at spatial scales below the resolution limit of conventional optical microscopy. To overcome this limit, various superresolution microscopy (SRM) techniques have been developed \cite{STED,MinFlux,PALM2006, Rust2006,Gustafsson2000}. However, higher resolution often comes at the cost of more complex optical setups \cite{Vangindertael_2018,SRM_demystified2019,SRM_brief_history2022,SRM_review_biological2022}.

One SRM technique that can be applied in a standard wide-field microscope without any modification is Super-resolution Optical Fluctuation Imaging (SOFI) \cite{SOFI2009}. SOFI exploits the temporal correlations of fluctuations of fluorescent labels to enhance the resolution. It has several advantages over other SRM techniques, such as no need for special instrumentation, lower sensitivity to emitter density than most single-molecule localization methods \cite{Geissbuehler2011}, and resolution improvement in three dimensions. Standard SOFI also has limitations, such as requiring a large number of frames and limited optical sectioning capability. The first limitation has recently been tackled by cleverly combining the temporal correlation analysis with deconvolution \cite{Zhao2023}. The second limitation has so far restricted the application of  SOFI to imaging modalities that do not suffer from out-of-focus background, such as Total Internal Reflection Fluorescence Microscopy. The usage of SOFI in thick samples is then limited to the sample surface -- 2D. A possible approach to realize 3D super-resolution inside thicker samples is to combine SOFI with light-sheet microscopy \cite{Light_sheetSOFI2020} or two-photon light-sheet for even deeper penetration \cite{chen_two-photon_2016}. Light-sheet microscopy, though, usually needs a dedicated setup and can't be incorporated into commercial wide-field microscopes. A different approach would be to modify the laser scanning confocal microscope (LSCM). The combination of SOFI with the additional modality of LSCM -- Image Scanning Microscopy -- offers 2.5 lateral resolution improvement \cite{Sroda2020} and can be a good choice for easy 3D super-resolution. However, as with all point scanning techniques, it is slow in acquiring larger volumes. 

Temporal Focusing (TF) is a method for fast, two-photon wide-field excitation providing sectioning in thick samples \cite{Oron_temporal, Zhu_temporal}. In this technique, a spectrally shaped pulse excites the fluorescent molecules only near the imaging plane. TF improves contrast in thick samples but also reduces the bleaching of out-of-focus dyes. TF can also be integrated in a standard wide-field microscope by using a femtosecond laser for excitation and a grating for pulse shaping.    
The simplicity made TF very versatile and it found interest in many areas in bio-imaging or neuroscience, namely: it is used in optogenetics and with uncaging \cite{Emiliani_review}, to observe cell dynamics \cite{CellulardynamicsTF2011}, to overcome scattering \cite{PTCSoLSTFScattering2018,descattering_single_pixel_patterns2018,PSo2021,DEEP1_2021}, and to achieve super-resolution \cite{Vaziri} or to obtain better contrast and sectioning \cite{Temporal_HiLo2,TSO_HiLo,axial_TF_SHepard_SO_2013}. TF was also used together with computational approaches \cite{DEEP2_2023,temporal_sieci}. The popularity and applicability of TF will probably be boosted by the wider availability of femtosecond fiber lasers that are stable and easy to use not only for physicists. 

In this work, we show that combining TF with SOFI is a promising approach for 3D imaging of thick samples. Our implementation provides optical sectioning in standard epi modality thanks to the two-photon excitation of a thin disc inside the specimen. This results in reduced bleaching, and better signal-to-background ratio. Our proposed setup is compact and simple, using a single diffraction grating and a robust and inexpensive fiber-based femtosecond laser for excitation.

\section{Materials and Methods}

\subsection{Super-resolution Optical Fluctuation Imaging}

Fluorescent markers commonly used in microscopy typically exhibit brightness fluctuations – blinking. In SOFI, blinking is a source of contrast rather than noise. Fluorescent emitters that are not interacting with each other blink independently, and as a result, brightness fluctuations carry information that can be used to break the diffraction limit. In SOFI, a series of frames (a movie of a fluctuating sample) is recorded and analyzed to form a single super-resolved image. If the emitter blinking is slow enough to be temporally resolved by a camera, the data recorded by each pixel contains a fluctuating intensity trace. In the simplest realization of SOFI, the variance of the fluctuating intensity at each pixel (2nd order cumulant) is used as a signal. This effectively shrinks the point spread function (PSF) of the detection system by a factor of $\sqrt{2}$ in all three dimensions \cite{SOFI2009}. In more advanced versions, higher-order cumulants are computed \cite{SOFI2009}. N-th order cumulant leads to the resolution improvement by a factor of $\sqrt{N}$, for a Gaussian detection PSF. However, higher-order cumulants need more frames and can produce cusp artifacts \cite{cusp2020}. As a result, typically only the 2nd order cumulants are used in practice.

Importantly, the computation of SOFI signal also eliminates non-fluctuating background, such as light coming from out-of-focus planes. This opens the possibility of 3D imaging with SOFI \cite{SOFI2009},  but 3D SOFI has some limitations, such as requiring a large number of frames and being prone to artifacts due to out-of-focus signals \cite{Pawlowska_2022}. Moreover, the bleaching of the emitters that are not in the imaging plane can decrease the quality of the 3D image or limit the imaging depth. The combination of SOFI with TF excitation reduces the out-of-focus signal and bleaching of out-of-focus emitters.

\subsection{Temporal Focusing}
TF \cite{Oron_temporal, Zhu_temporal} is a technique for wide-field two-photon excitation that provides sectioning -- excitation of an axially confined volume of the sample. TF's sectioning is caused by the change of pulse duration with its propagation along the optical axis. As the rate of two-photon excitation depends on the instantaneous intensity raised to the second power, the shorter the pulse the more molecules are excited and emit fluorescence. 

To change the pulse duration upon propagation one can use dispersion. While propagating, the pulse's spectral components can acquire different phases. When there is a mismatch in the respective phases of these spectral components, the pulse elongates.  

Dispersion can be then caused by introducing an optical path difference for the spectral components of a pulse. Such geometrical dispersion is used in TF, where a diffraction grating makes the spectral components travel in slightly different directions. To focus the pulse in time the optical paths for every spectral component should be identical. This  situation happens if we image the surface of the grating using a 4f telescope consisting of a L3 lens, and an objective lens (see scheme \ref{scheme}a). 

After calculating the geometric dispersion we can approximately calculate the pulse duration and two-photon fluorescence rate in the vicinity of temporal focus. We can model the sectioning capability of TF using the following equation \cite{Dana_2011_intensityeqn}:
\begin{equation}
    F(z) \propto \frac{1}{\sqrt{1+a(z-z_0)^2}}, 
\end{equation}
where $F(z)$ is the two-photon signal from a given z-plane, and "$a$" describes the sectioning capability of TF. $z_0$ is the axial position of the temporal focus. The sectioning capability "$a$" depends on the pulse duration, the grating line density, the 4f telescope magnification, the refractive index of the sample, and the central wavelength of the excitation. If we fill the whole entrance pupil of the microscope objective with spectral components, the sectioning capability of TF is restricted by the numerical aperture (NA) of the objective and the excitation wavelength. The axial position of the temporal focus can be changed by adjusting the initial chirp of the pulse (an option in femtosecond lasers to cancel the pulse broadening caused by optical elements) or by changing the distance between the grating and the first lens of the telescope.

\subsection{Experimental Setup and Measurement Procedure} 

\begin{figure}[hb]
\centering\includegraphics[width=10cm]{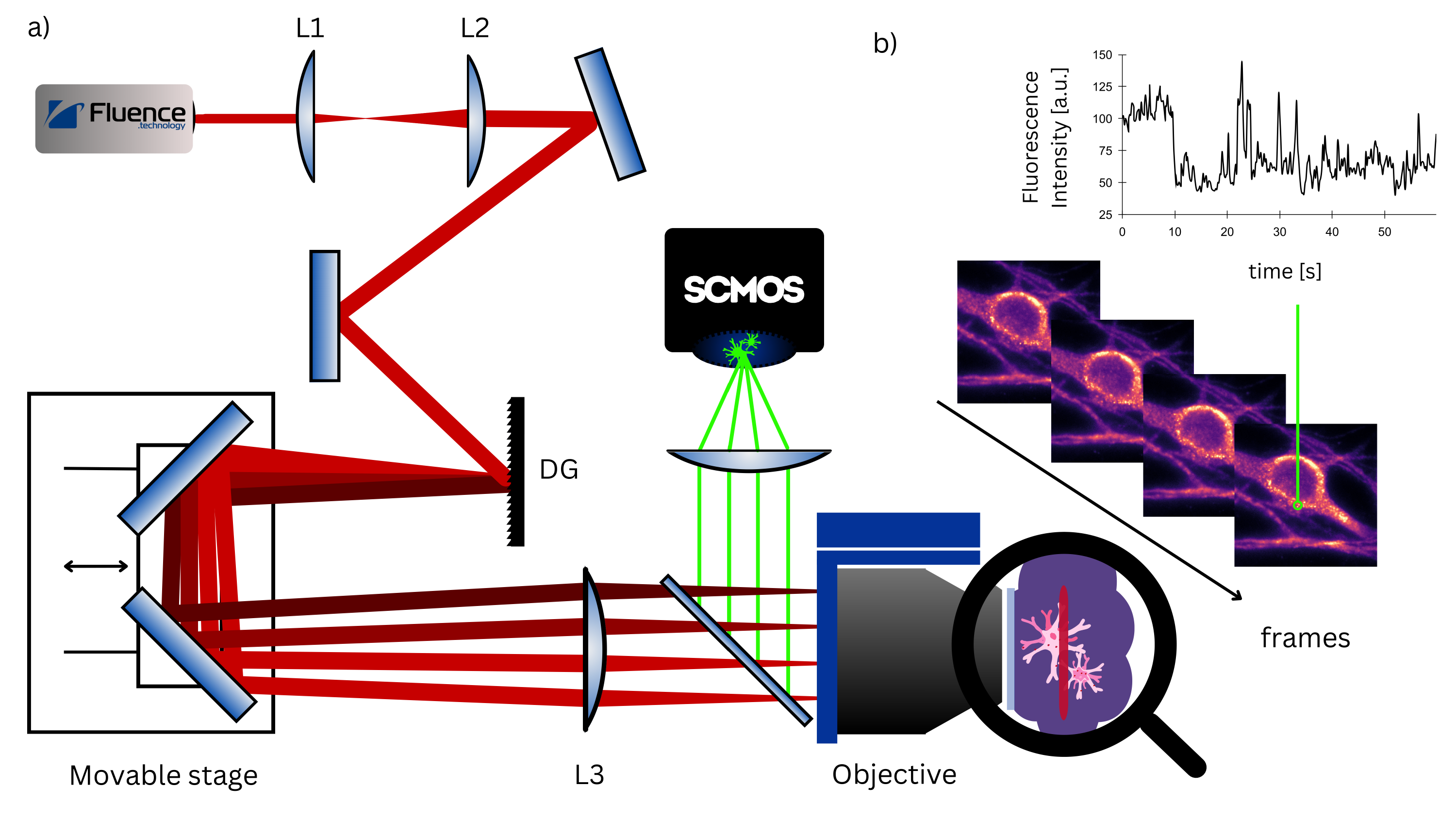}
\caption{Schematic of the temporal focusing setup combined with the SOFI analysis. In part a), a femtosecond laser beam, magnified by a telescope composed of L1 ($f_1=75$ mm) and L2 ($f_2=500$ mm), is directed onto a DG – a blazed grating with 600 lines/mm. A movable stage is used to align the temporal focus with 
an objective focus. The L3 and the microscope objective form a 4f telescope. A dichroic mirror reflects the sample’s fluorescence to a camera. Part b) displays a sequence of frames captured by the camera, where the signal at a single pixel varies over time due to blinking. The frames are subsequently subjected to SOFI analysis.}


\label{scheme}
\end{figure}

The schematic representation of the experiment is shown in Figure \ref{scheme}. As the excitation source, we use a femtosecond fiber laser (Fluence Halite 2). The laser has a central wavelength of 1030 nm, and 20 nm bandwidth that results in a minimum pulse duration of around 190 fs full width at half maximum (FWHM). We use a mean power of 2W at 20MHz repetition rate to achieve a field of view of around 40 $\mu$m diameter. For this, we enlarge the beam to a diameter of $8$ mm with a telescope consisting of two lenses L1 ($f_{1}=75$ mm) and L2 ($f_{2}=500$ mm). With the enlarged beam, we illuminate a blazed grating DG (600 lines/mm). The grating is perpendicular to the central wavelength’s first order of diffraction. We use two mirrors on a translation stage to reflect the diffracted light. By moving the stage with the mirrors, we can adjust the axial position of the temporal focus. The position of these mirrors sets the optical path between the grating DG and the next lens L3 ($f_{3}=$400 mm). We selected $f_3$ such that the entire 20 nm bandwidth of the excitation pulse matches the objective’s pupil size. Finally, we use an oil immersive objective (Nikon PlanApo, NA = 1.4, WD = 0.13, magnification = 100x) to illuminate the sample. The objective and L3 lens form a 4f imaging setup that demagnifies the excitation beam by a factor of 200, resulting in a diameter on the sample of around 40 $\mu$m that determines the size of the field of view. The field of view is restricted by the size of the mirrors and lenses (1 inch) and the available peak power of the excitation beam. 

To image the two-photon excited fluorescence, we separate it from the excitation light using a dichroic mirror (Thorlabs DMS750) and two short pass filters (Thorlabs FESH0750). A 200 mm tube lens (Thorlabs TTL200-A) focuses the fluorescence onto the sensor of QCMOS camera (Hamamatsu Orca-Quest). The highest Signal-to-Noise-Ratio for this camera is achieved for the photon number resolving mode, which has a dynamic range of 200 counts and a readout speed of 5 frames per second. 

We acquired 3D volume image data by adjusting the objective axial position with a z-focusing piezo stage (MadCityLabs). For each z-position, we recorded 300 frames with an exposure time of 1 ms. The short exposure and long readout results from the photon number resolving mode.

We analyze the recorded movies with a Matlab script using the open-source Localizer library \cite{localizer} for computing second-order SOFI. We also used SOFI Evaluator \cite{moeyaert_sofievaluator_2020} to confirm that the computed contrast comes from optical fluctuations and not other time-dependent effects such as bleaching.

\section{Results}

\begin{figure}[hb]
\centering\includegraphics[width=10cm]{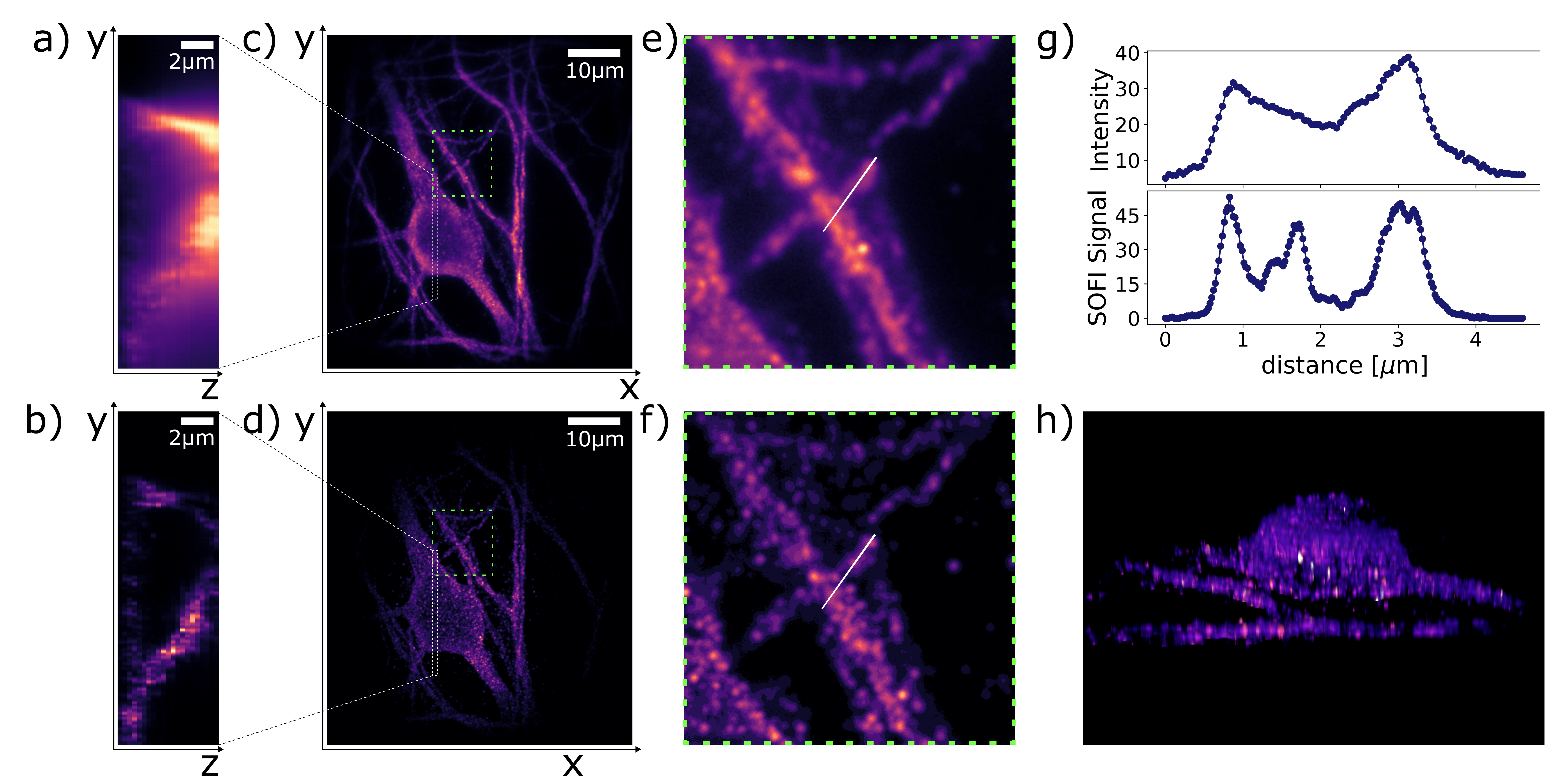}
\caption{Comparison of the mean intensity images and the SOFI images of rat neurons stained with anti-MAP2 qdot-conjugated antibody obtained with wide-field two-photon excitation with Temporal Focusing. (a) and (b) depict YZ cross-sections of the mean intensity and SOFI images, respectively. (c) and (d) display the full field of view of the mean intensity and SOFI images, respectively, revealing the soma and dendrites. (e) and (f) show magnified areas of interest from (c) and (d), respectively, emphasizing dendritic features. Plot (g) represents a signal along the white lines indicated in (e) and (f), illustrating the enhanced contrast and improved lateral resolution of SOFI compared to mean intensity. (h) is a 3D rendering of a neuron from SOFI images, showcasing SOFI’s ability to reconstruct 3D morphology. }
\label{result}
\end{figure}

For this study, we stained a cell culture of rat hippocampal neurons with anti-Microtubule Associated Protein 2 (MAP2) -- a quantum dots-conjugated antibody. The staining procedure was based on \cite{prost_working_2016} and is described in the supplementary material 1.1. A field of view of 40 $\mu$m allowed us to image a whole cell soma. As shown in Figure (\ref{result}c), we could see the individual neural cell with dendrites. In order to evaluate 3D performance of our technique we collected z-stack with 21 "z" positions separated by 300 nm. For every "z" position we took 300 frames and performed SOFI analysis (bottom row of the Figure \ref{result}). SOFI removes nonfluctuating background, and as a result improves z-sectioning. This improvement can be clearly seen in comparison between the mean intensity yz-cross-section (\ref{result}a) and SOFI signal yz-cross-section (\ref{result}b). 
Another important factor is the lateral super-resolution of SOFI.   The theoretical FWHM of PSF for the intensity image was 255 nm, while we measured $274 \pm 17$ nm. For the SOFI images, the experimental PSF was $199 \pm 12$ nm, consistent with the $\sqrt{2}$ theoretical resolution improvement. The close-up views of dendrites (\ref{result}e) -- mean intensity and (\ref{result}f) -- SOFI  present that the SOFI image has better contrast and better resolution. Such an improvement enables to distinguish more features of dendrites. The signal profile  (\ref{result}g) along white lines marked on (\ref{result}e) and (\ref{result}f) shows that SOFI has better contrast and also makes possible to resolve two emitters that are not visible in the mean intensity picture. Better resolution, higher contrast and slow bleaching of out-of-focus dyes makes TF and SOFI a great synergy for 3D imaging -- see Figure (\ref{result}h), a  3D rendering of the observed structure.

To evaluate 3D sectioning capability of  TF and SOFI, we used a flat sample of spin-coated quantum dots (QDot™ 605 ITK™, Q21701MP, Invitrogen™) dissolved in PMMA. We scanned the sample through the temporal focus with 50 nm steps and for every position we recorded a movie from which we computed SOFI images. We summed the signal from pixels for every SOFI image corresponding  to a "z" position. From obtained plots we calculated that the sectioning for SOFI was $481\pm14$ nm FWHM. The result was much better than for the TF sectioning alone ($2.45\pm0.23$ $\mu$m FWHM). Interestingly, we obtained the same sectioning values for SOFI with TF and with wide-field two-photon excitation, which means that TF does not enhance SOFI's sectioning compared to wide-field excitation. However, wide-field excitation produces an out-of-focus signal that distorts the SOFI image and causes unnecessary bleaching. 


\section{Conclusions} 

To conclude, we have demonstrated the possibility of combining Temporal Focusing with SOFI. This combination results in a moderately fast 3D super-resolution technique available in epi-geometry. We achieved 200 nm lateral resolution and 480 nm axial resolution in the whole volume of the sample. Such a feature could be obtained by slightly modifying a commercial microscope as both methods (TF and SOFI) are quite easy to implement. 

Our technique's speed is limited mostly by the equipment we used (available laser power, camera frame rate). In principle, SOFI should work with the timescales of seconds for a 3D Super-resolution image, and with recent advances \cite{Zhao2023} it could be even faster. The new computational approach is complementary to our idea as TF is a wide-field excitation method with the speed restricted only by the camera's frame rate. 

In this work, we used quantum dots, as they are photostable and blink with suitable timescales, but TF SOFI could also work with proteins such as Skylan-S \cite{blinking_proteins} or Dronpa \cite{Proteins_Dedecker}. Since we use the blinking property to improve the resolution, the blinking statistics itself can contain information about the micro-environment of the sample \cite{Geissbuehler2012}. TF SOFI enables access to this information in 3D volume. 

Another possible application for our technique is to improve resolution for thick scattering samples. Adding a line scan to TF \cite{Oron_linescan} would increase sectioning, reduce out-of-focus bleaching, and decrease the excitation power needed for a large field of view. The scanning line can also be synchronized with the camera's rolling shutter to act as a digital pinhole. Such a realization will reject scattered light and could be a reliable technique for 3D super-resolution deep imaging. SOFI should also work with scattering cases as we expect that the scattered light, similar to the out-of-focus background, is uncorrelated. Additionally, it would be interesting to combine TF with computational methods and patterned illumination \cite{DEEP1_2021,DEEP2_2023} to overcome scattering and then connect with SOFI. 

In summary, TF SOFI can be easily implemented in standard fluorescence microscopes and offers fast and robust 3D super-resolution imaging suitable for structural imaging, deep imaging, and even live imaging.

\setcounter{section}{0}
\section{Supplementary Material}

\subsection{Staining procedure}
SOFI requires fluorescent labels that blink at a rate matched to the camera frame rate. We use quantum dots as the labels because they have high brightness, photostability, and suitable blinking properties. We follow the immunocytochemistry staining protocol adapted from \cite{prost_working_2016}, where the authors found that the fluorescence of some quantum dot conjugated antibodies fades quickly with common buffers. They suggested using a background-reducing agent BKRA (DAKO, S3022) to prevent quenching.
The steps of our adapted protocol are as follows:
\begin{itemize}
    \item  fix three weeks old rat primary dissociated mixed culture with 4\% PFA and 4\% sucrose.
    \item  permeabilize the culture with 0.1\% triton for 10 minutes,
    \item   rinse the culture three times with PBS,
    \item  incubate the culture in 10\% normal donkey serum (Sigma-Aldrich\textsuperscript{\textregistered}, S30) for blocking (1.5h),
    \item stain the culture with Microtubule Associated Protein 2 (MAP2) antibody diluted in PBS with 2\% normal donkey serum overnight at 4 deg C,
    \item wash the culture in BKRA solution 1/50 in 50mM Tris, as recommended in \cite{prost_working_2016},
    \item incubate the culture with Donkey anti-Mouse Secondary Antibody, Qdot625 (Invitrogen™ Q22085) diluted 1:300 in BKRA for 1h at room temperature,
    \item rinse the culture three times in BKRA solution (1/50 in 50mM Tris),
    \item mount the coverslip with ProLong™ Glass Antifade Mountant with NucBlue™ (Invitrogen™, P36981).
\end{itemize}
This staining procedure allowed us to perform experiments for several days after the staining. In the first days, we observed the blinking quantum dots that had a granular structure. After about two weeks, the granular structure disappeared, the fluorescence weakened and hardly any SOFI signal could be detected.

\begin{backmatter}
\bmsection{Funding}
This work was supported by the Foundation for Polish Science under the FIRST TEAM project “Spatiotemporal photon correlation measurements for quantum metrology and super-resolution microscopy” co-financed by the European Union under the European Regional Development Fund (POIR.04.04.00-00-3004/17-00), by the National Science Centre, Poland, grant number 2022/47/B/ST7/03465, and Horizon Europe MSCA FLORIN project ID 101086142.
\bmsection{Acknowledgments}
We want to express our gratitude to Aleksandra Mielnicka who helped us in sample preparation. Thanks to Wim Vandenberg for consultations about desire field-of-view, and for Aleksander Krupiński-Ptaszek for helping with data analysis and interpretation. We particularly appreciate very deep and precise comments from Adrian Makowski who helped us in revision of this manuscript. \newline
Latex template adopted from Optica latex template.

\bmsection{Disclosures}
The authors declare no conflicts of interest.

\bmsection{Data availability} Data underlying the results of sectioning presented in this paper is available in Ref. \cite{data}.

\end{backmatter}


\bibliography{sample}






\end{document}